\documentclass[preprint]{imsart}
\newcommand\independent{\protect\mathpalette{\protect\independenT}{\perp}}
\def\independenT#1#2{\mathrel{\rlap{$#1#2$}\mkern5mu{#1#2}}}

\newcommand{\E}{\mathbb{E}}
\newcommand{\Prob}{\mathbb{P}}

\usepackage[figuresright]{rotating}
\usepackage{amsmath}
\usepackage{amssymb}
\usepackage{subfigure}
\usepackage{pstricks,pst-node,pst-plot}
\usepackage{color}
\usepackage{pst-all}
\usepackage{pspicture}
\usepackage{booktabs}
\usepackage{wasysym}
\usepackage[all]{xy}
\xyoption{arc}
\usepackage{epsfig,graphicx}
\usepackage{theorem}
\usepackage{natbib}

\newcounter{count}

\newenvironment{proposition}[1][Proposition \arabic{count}]{\vspace{1em}\refstepcounter{count}\begin{trivlist}
\item[\hskip \labelsep {\bfseries #1}]\em }{\end{trivlist}\vspace{1em}}

\newenvironment{proof}[1][Proof]{\vspace{1em}\refstepcounter{count}\begin{trivlist}
\item[\hskip \labelsep {\bfseries #1}]}{\hfill$\Box$\end{trivlist}\vspace{1em}}

\newenvironment{definition}[1][Definition \arabic{count}]{\vspace{1em}
\refstepcounter{count}\begin{trivlist}
\item[\hskip \labelsep {\bfseries #1}]}{\end{trivlist}\vspace{1em}}

\begin{document}
\setcounter{count}{0}
\begin{frontmatter}

\title{Intervention in undirected Ising graphs and the partition function}
\runtitle{Intervention in undirected graphs}

\author{Lourens Waldorp}
\author{Maarten Marsman}
\runauthor{Waldorp and Marsman}

\address{\tt\small waldorp@uva.nl\hspace{3em} m.marsman@uva.nl}
\begin{abstract}
Undirected graphical models have many applications in such areas as machine learning, image processing, and, recently, psychology. Psychopathology in particular has received a lot of attention, where symptoms of disorders are assumed to influence each other. One of the most relevant questions practically is on which symptom (node) to intervene to have the most impact. Interventions in undirected graphical models is equal to conditioning, and so we have available the machinery with the Ising model to determine the best strategy to intervene. In order to perform such calculations the partition function is required, which is computationally difficult. Here we use a Curie-Weiss approach to approximate the partition function in applications of interventions. We show that when the connection weights in the graph are equal within each clique then we obtain exactly the correct partition function. And if the weights vary according to a sub-Gaussian distribution, then the approximation is exponentially close to the correct one. We confirm these results with simulations. 
\end{abstract}

\end{frontmatter}

\section{Introduction}\noindent
Graphical models are popular in many applications such as machine learning, image processing, social science, and recently psychology. 
One of the earlier applications was in expert systems, where the objective was to determine the probability of a correct diagnosis given a specific configuration of symptoms and screenings \citep{Cowell:1999}. Such applications are also extremely relevant to psychology. Paramount to expert systems the effect of interventions (medication or therapy) is of interest. \citet{Lauritzen:2002} showed that intervention by replacement (hard intervention) in undirected graphs is equivalent to conditioning, unlike intervention in directed acyclic graphs. As a consequence, no special treatment to determine (marginal) probabilities is required for interventions in undirected graphs. 

Specifically, for the Ising model, where binary nodes are modeled by their values and their interactions with neighbouring (i.e., connected) nodes, determining the probability of an intervention means that we simply fix a variable to a specific value (0 or 1) and then determine the probabilities as we would in the conditional distribution. We do however require the partition function (normalising constant) which It boils down to marginalising with specific values for the conditioning variables. For the Ising model the marginal is computationally intensive, with $2^{k}$ elements  to consider in the subset of nodes $V_{k}$. Approximations to the partition function can be used to obtain approximate probabilities. One approach is to ignore the interactions altogether, which simplifies the partition function to a product of the partition function for each node separately. Another approach is to obtain upper and lower bounds of the partition function, like the Bethe lattice \citep{Wainwright:2008} or the related version of locally tree-like graphs \citep{Dembo:2013}. Here use a different approach and consider the fact that each clique is in itself a Curie-Weiss model (a fully connected graph), for which the partition function can be determined in linear time $O(|\mathcal{C}|k)$, where $\mathcal{C}$ is the set of all cliques and  $k$ is the size of the largest clique. In a Curie-Weiss model the edge weights are considered equal, which is obviously inappropriate in many situations. We therefore determine that the error of approximation when the variation in edge weights is limited to sub-Gaussian random variables is $O_{p}(k^{-1/2})$, with $k$ the size of the clique. 

We first discuss undirected graphical models in Section \ref{sec:graphical-models}. Next we discuss how interventions can be defined on undirected graphical models in Section \ref{sec:intervention} and how such conditioning is implemented in Ising models. Here the problem is the normalising constant (partition function) that prohibits calculation of probabilities. In Section \ref{sec:normalising-constant} we discuss possible solutions and present our approach based on the Curie-Weiss model. In Section \ref{sec:numerical-illustration} we perform several simulations to illustrate the size of the errors in the normalising constant with the Curie-Weiss model. Proofs can be found in the Appendix.

\section{Undirected graphical models}\label{sec:graphical-models}\noindent
An undirected graphical model or Markov random field is a set of probability distributions representing the structure of some graph $G$.  There are two equivalent ways of defining a Markov random field: (i) in terms of Markov properties and (ii) in terms of the factorization property. 

Let $G=(V,E)$ be an undirected graph, where $V$ is the set of nodes  $\{1,2,\ldots,n\}$ and $E=V\times V$ is the set of edges $\{(s,t): s,t\in V\}$, with size $|E|=m$. A subset of nodes $Q$ is a cutset or separator set of the graph if removing $Q$ results in two (or more) components. For instance, $Q$ is a cutset if any path between any two nodes $s\in A$ and $t\in B$ must go through some $q\in Q$. A clique is is a subset of nodes in $C\subset V$ such that all nodes in $C$ are connected, that is, for any $s,t\in C$ it holds that $(s,t)\in E$. A maximal clique is a clique such that including any other node in $V$ will not be a clique. 

For an undirected graph $G$, we associate with each vertex $s\in V$ a random variable $X_{s}\in\mathcal{X}$. For any subset $A\subset V$ of nodes we define a configuration $x_{A}=\{x_{s}: s\in A\}$. 
A configuration $x_{C}$ for $C\subset V$ is $\{x_{s}, s\in C\}$. An edge set restricted to the edges among a subset $D\subseteq V$ is denoted by $E_{D}$. 

Two variables $X_{s}$ and $X_{t}$ are independent if $\Prob(X_{s},X_{t})=\Prob(X_{s})\Prob(X_{t})$, and we write this as $X_{s}\independent X_{t}$. The variables $X_{s}$ and $X_{t}$ are conditionally independent on $X_{j}$ if $\Prob(X_{s},X_{t}\mid X_{j})=\Prob(X_{s}\mid X_{j})\Prob(X_{t}\mid X_{j})$.
For subsets of nodes $A$, $B$, and $W$, we denote by $X_{A}\independent X_{B}\mid X_{W}$ that $X_{A}$ is conditionally independent of $X_{B}$ given $X_{W}$. 
A random vector $X$ is {\em Markov compatible} or {\em Markov} with respect to $G$ if $X_{A}\independent X_{B}\mid X_{W}$ whenever $W$ is a cutset that yields two disjoint subsets $A$ and $B$.
For strictly positive distributions the Hammersly-Clifford theorem says that the Markov property is equivalent to the factorisation property \citep{Cowell:1999,Lauritzen96}. The distribution of the random vector $X$ is said to {\em factorise according to graph} $G$ if it can be represented by a product of compatibility functions (not necessarily probabilities in general) of the cliques
\begin{align}\label{eq:factorisation}
p(x) = \frac{1}{Z_{V}}\prod_{C\in \mathcal{C}} \psi_{C}(x_{C})
\end{align}
where $\psi_{C}$ are compatibility functions for clique $C$. This factorisation is convenient since it implies that the effects of conditioning can be evaluated for each clique separately. 

One of the most well known binary undirected graphical models is the Ising model, known from statistical physics to model the magnetic field \citep[see e.g., ][]{Kindermann:1980,Cipra:1987,Kolaczyk:2009}. The Ising model considers cliques of sizes one and two nodes only, so the interactions are at most pairwise \citep{Wainwright:2008,Besag:1974}. Let $\theta$ be the parameter vector containing all parameters. The distribution of {\em the Ising model} can be written as
\begin{align} 
p_{\theta}(x) = \exp\left(\sum_{s\in V} \theta_{s} x_{s} + \sum_{(s,t)\in E} \theta_{st}x_{s}x_{t} - A(\theta)\right)
\label{eq:ising}
\end{align}
where 
\begin{align*}
A(\theta) :=\log \sum_{x\in \{0,1\}^{p}} \exp\left(\sum_{s\in V} \theta_{s} x_{s} + \sum_{(s,t)\in E} \theta_{st}x_{s}x_{t} \right)
\end{align*}
is the log normalization constant. It is immediate that the Ising model is exponential family with sufficient statistics $\phi(x)=(x_s, s\in V; x_s x_t, (s,t)\in E)$. It is also minimal since the functions in $\phi(x)$ are linearly independent, i.e. $\langle u,\phi(x)\rangle$ is not a constant a.e. for any nonzero $u\in \mathbb{R}^{|V|+|E|}$.

\section{Intervention and conditioning graph}\label{sec:intervention}\noindent
The general idea of an intervention graph is the same as the causal directed graph \citep{Lauritzen:2001}. An intervention is defined as a manipulation from outside the graph such that a variable (or set of variables) is fixed (clamped) to a particular value, where no other variable can affect this conditioning node \citep{Spirtes:1996,Eberhardt:2007}. This is equivalent to the do-operation \citep{Pearl:2000}. No other nodes are affected directly by the intervention except those in the conditioning set.  In an undirected graph the clique structure remains the same and the values are replaced by $x_{i}^{\star}$ (intervention by replacement). We then want that the factorisation of the graph remains and the conditioning on the nodes in the cliques $C\in \mathcal{C}$ that intersect with the intervention nodes $A$, does not disrupt the factorisiation (i.e., the graph remains Markov compatible). This leads to the following definition of intervening in undirected graphs \citep{Lauritzen:2001,Lauritzen:2002}. 
%
\begin{definition}{ (Causal undirected graph)}\label{def:causal-undirected-graph}\label{def:causal-undirected-graph}
Let $G$ be a graph with Markov compatible distribution $p(x) = \prod_{C\in \mathcal{C}} p_{C}(x_{C})$ 
over the clique set $\mathcal{C}$ in $G$. Furthermore, let $x_{A}^{\star}$ be the values of the nodes in the subset $A$ that replace the original values. Then we call it a {\em causal undirected graph for} $A\subseteq V$ if 
\begin{align}\label{eq:causal-undirected-graph}
p(x \mid\mid x_{A}^{\star}) =  \prod_{C\in \mathcal{C}} p_{C}( x_{C\backslash A} \mid\mid x_{C\cap A}^{\star})
\end{align}
\end{definition}
Note that when $C\cap A=\varnothing$, then there is no intervention. We can equivalently write 
\begin{align}\label{eq:causal-undirected-graph-2}
p(x\mid\mid x_{A}^{\star}) =  \prod_{\mathcal{C}\ni C\cap A\ne \varnothing} p_{C}( x_{C\backslash A} \mid\mid x_{C\cap A}^{\star})\prod_{\mathcal{C}\ni C\cap A= \varnothing} p_{C}( x_{C})
\end{align}
where $\mathcal{C}\ni C\cap A\ne \varnothing$ identifies sets $C\cap A\ne \varnothing$ with $C$ in $\mathcal{C}$. We see from this definition that we need only determine the intervention locally, with respect to the clique. Suppose that we intervene on node $j$ in clique $C\in \mathcal{C}$ with the value $x_{j}^{\star}$. Then we have from our definition that we only need the cliques $C\in \mathcal{C}$ such that $j\in C$ to obtain $
p_{C}(x_{C\backslash j}\mid\mid x_{j}^{\star})$ and the rest of the terms in the factorisation remain as before. 

The definition is still unclear on what for each clique factor $p_{C}( x_{C\backslash A} \mid\mid x_{C\cap A}^{\star})$ means. \citet{Lauritzen:2002} show that in undirected graphs with finite state space for each $x_{i}$ intervening by replacement (do-operation) is equivalent to conditioning, that is
\begin{align}\label{eq:intervention-conditioning}
p_{C}( x_{C} \mid\mid x_{C\cap A}^{\star}) = p_{C}( x_{C\backslash A} \mid x_{C\cap A}^{\star})
\end{align}
The reason is that the structure of the undirected graph is not changed when intervening, at least not when using intervention by replacement. For directed (acyclic) graphs this is different because any incoming edges (arrows) on the intervention nodes will be deleted since the intervention completely controls them, and no other variables can affect the intervention nodes \citep{SpirtesGlymourScheines93,Lauritzen:2001}. This changes the structure of the graph and therefore changes the distribution, and this difference between intervention and conditioning can be detected. In undirected graphs we cannot distinguish between having observed or intervened on values in $x_{C\cap A}^{\star}$. In an undirected graph nothing of the structure is changed and so there is no difference between intervention or conditioning to be detected in terms of conditional independencies.

Equivalently, we can think of an intervention on node $j$ as an additional node $I_{j}$ in the set $\{\text{on},\text{off}\}$ directly connected to the intervention node $j$ in the intervention set $A$ \citep{SpirtesGlymourScheines93, Eberhardt:2007}. This node $I_{j}$ sets node $j$ to on or off (do-operation) and only node $j\in A$. If $I_{j}$ sets the node to off, then the observational distribution with respect to node $j$ obtains. If $I_{j}$ sets node $j$ to on, then the structure in $G$ remains unchanged, resulting in the same factorisation as without intervention but with the value $x_{j}$ of node $j$ set to 0 or 1. For each node in the intervention set $A$ there is a node $I_{i}$ that is connected only directly to node $i\in A$; collectively such exogenous nodes are referred to as $I_{A}$, where each node in $A$ is connected to a single node $I_{i}$ with $i\in A$.

Consider the graph $G=(V,E)$ in Figure \ref{fig:cond-vs-do}(a) with five nodes and let $x\in \{0,1\}^{5}$ be a binary vector. There are three cliques, $\{1,5\}$, $\{1,2\}$, and $\{2,3,4\}$. The joint distribution is
\begin{align}
p(x) = p(x_{1},x_{5})p(x_{1},x_{2})p(x_{2},x_{3},x_{4})
\end{align}
where we used the factorisation in (\ref{eq:factorisation}). According to our definition of intervention, an intervention on node $2$ would result in the distribution
\begin{align*}
p(x \mid\mid x_{2}^{\star}) = p(x_{1},x_{5})p(x_{1}\mid\mid x_{2}^{\star})p(x_{3},x_{4}\mid\mid x_{2}^{\star})
\end{align*}
But by the fact that the intervention distribution equals the conditional distribution without $A=\{2\}$, we obtain 
\begin{align*}\label{eq:intervention}
p(x \mid\mid x_{2}^{\star}) = p(x_{1},x_{5})p(x_{1}\mid x_{2}^{\star})p(x_{3},x_{4}\mid x_{2}^{\star})=p(x_{\backslash 2}\mid x_{2}^{\star})
\end{align*}
where $x_{\backslash 2}:=x_{V\backslash \{2\}}$. And we observe that conditioning obtains the same distribution as intervening in undirected graphs. 
\begin{figure}[t]
\begin{minipage}[b]{0.3\textwidth}\centering
\begin{align*}
\xymatrix{
*++[o][F]{1}\ar@{}[r]_>{\displaystyle C_{1}}&{\xy(-19,-20);(-6,5) **\frm<44pt>{--}\endxy}	&*++[o][F]{3}\ar@{-}[rr]\ar@{-}[d]\ar@{-}[r]	&&*++[o][F]{4}\\
*++[o][F]{5}\ar@{-}[u]^{\displaystyle C_{2}} 	&{\xy(4,-5);(45,20) **\frm<44pt>{--}\endxy} &*++[o][F]{2}\ar@{-}[urr]\ar@{-}[ull] &
\xy 
/r6pc/:p-(.85,-0.32),
{\ellipse(,.31){--}}
\endxy
&C_{3}
}
\end{align*}
(a)
\end{minipage}
\hspace{1.5em}
\begin{minipage}[b]{0.3\textwidth}\centering
\begin{align*}
\xymatrix{
*++[o][F]{1}\ar@{}[r]			&	&*++[o][F]{3}\ar@{-}[rr]\ar@{-}[d]\ar@{-}[r]	&&*++[o][F]{4}\\
*++[o][F]{5}\ar@{-}[u]		&	&*++[o][F**:pink]{2}\ar@{-}[urr]\ar@{-}[ull]&&*++[o][F]{I_{2}}\ar@{->}[ll]
}
\end{align*}
(b)
\end{minipage}
\setcounter{figure}{0}
\caption{Graph of 5 nodes with cliques $C_{1}=\{1,2\}$, $C_{2}=\{1,5\}$ and $C_{3}=\{2,3,4\}$ represented in (a). In (b) equivalently intervening (conditioning) on node 2 in an undirected graph by an auxiliary variable $I_{2}$ that determines the output of node 2. }
\label{fig:cond-vs-do}
\end{figure}
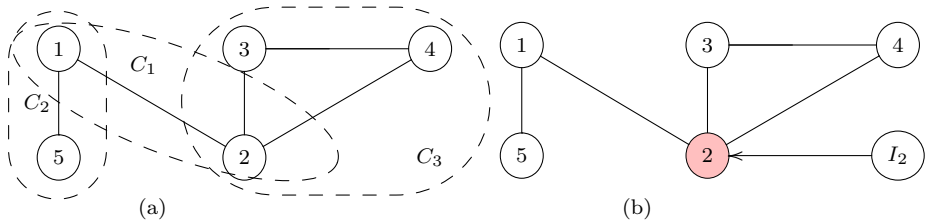

A representation of the equivalent version of intervening by an exogenous node $I_{j}$ is shown in Figure \ref{fig:cond-vs-do}(b). It has the same setup as the previous example shown in Figure \ref{fig:cond-vs-do}(a) but now the exogenous variable $I_{2}$ is added. It is clear that the same node $I_{2}$ cannot simultaneously intervene on another node as this would in general lead to possible spurious connections between the endogenous nodes in $V$. 

\section{Conditioning in Ising models}\label{sec:ising-models}\noindent
In the Ising model any edge $(s,t)\in E$ is represented by the product $\theta_{st}x_{s}x_{t}$, and there are no higher order terms. Consider again Figure \ref{fig:cond-vs-do} with five nodes. 
If we assume that the external field is 0, then we only have the products of the cliques from the factorisation. So the joint distribution for the Ising model for Figure \ref{fig:cond-vs-do} can be written as
\begin{align} 
p_{\theta}(x) &= \frac{1}{Z(\theta)}\exp\left( \theta_{15}x_{1}x_{5}\right)\exp\left(\theta_{12}x_{1}x_{2}\right)\exp\left( \theta_{23}x_{2}x_{3} + \theta_{24}x_{2}x_{4} + \theta_{34}x_{3}x_{4}\right)
\end{align}
where $Z(\theta)=\exp(A(\theta))$ is the normalising constant. And we immediately see we have the factorisation in (\ref{eq:factorisation}). Then conditioning on node 2 having value $x_{2}^{\star}$ requires we consider the normalising constant $Z_{\backslash 2}$ over the remaining variables $V\backslash \{2\}$. From the factorisation and equality of intervening and conditioning in (\ref{eq:intervention}), we have that we can consider each clique separately and then plug in the value $x_{2}^{\star}$ for conditioning. So, for the clique $\{1,2\}$ we get
\begin{align*}
p(x_{1}\mid x_{2}^{\star}) = \frac{\exp(\theta_{12}x_{1}x_{2}^{\star})}{1 + \exp(\theta_{12}x_{2}^{\star})}
\end{align*}
The normalising constant in the denominator is obtained by plugging in the possible values 0 and 1 respectively, obtaining $\exp(\theta_{12}0x_{2}^{\star})+\exp(\theta_{12}1x_{2}^{\star})$; we denote this normalising constant by $Z_{1,\backslash 2}$. And for the clique $\{2,3,4\}$ we obtain
\begin{align*}
p(x_{3},x_{4}\mid x_{2}^{\star}) = \frac{\exp\left( \theta_{23}x_{2}^{\star}x_{3} + \theta_{24}x_{2}^{\star}x_{4} + \theta_{34}x_{3}x_{4}\right)}{1 + \exp(\theta_{23}x_{2}^{\star})+ \exp(\theta_{24}x_{2}^{\star}) + \exp(\theta_{34}) }
\end{align*}
where the normalising constant $Z_{34,\backslash 2}$ of the denominator is determined by the values of $(x_{3},x_{4})$ with $(0,0)$, $(0,1)$, $(1,0)$, and $(1,1)$. For the clique $\{1,5\}$ we need not change anything, and so remains $p(x_{1},x_{5})$. We then obtain the conditional distribution from the factorisation theorem by making the product $p(x_{1},x_{5})p(x_{1}\mid x_{2}^{\star})p(x_{3},x_{4}\mid x_{2}^{\star})$
\begin{align} 
p_{\theta}(x_{\backslash 2}\mid x_{2}^{\star}) &= \frac{1}{Z_{\backslash 2}}\exp\left( \theta_{15}x_{1}x_{5}\right)\exp\left(\theta_{12}x_{1}x_{2}^{\star}\right)\exp\left( \theta_{23}x_{2}^{\star}x_{3} + \theta_{24}x_{2}^{\star}x_{4} + \theta_{34}x_{3}x_{4}\right)
\end{align}
where 
\begin{align*}
Z_{\backslash 2} = Z_{15,\backslash 2}Z_{1,\backslash 2}Z_{34,\backslash 2}
\end{align*}
Note that the complexity of the normalising constant depends on the size of the clique because the different cliques are conditionally independent; the larger the clique the more complex it will be to determine. In fact the complexity is $O(2^{k})$ with $k$ the number of nodes in the clique, and so is exponential in the number of nodes in a clique \citep{Wainwright:2008}. For small graphs the complexity is not prohibitive, since it can be computed directly. However, the complexity is problematic for large graphs where the clique size is large. In such cases we require a computationally efficient way to obtain the normalising constant of the clique, save the nodes that are conditioned on.

\section{Determining the normalising constant}\label{sec:normalising-constant}
From the last example in the previous section it is clear that the normalising constant for the conditional distribution can be cumbersome but, fortunately, it can be factorised. The normalising constant is in general for the Ising model an NP-hard problem \citep[see, e.g., ][]{Wainwright:2008}. For instance, with 30 nodes we already require a sum over more than a billion terms. In optimisation algorithms the calculation of the normalising constant typically has to be calculated thousands of times, and so this makes such calculations infeasible. 

A naive approach would be to simply ignore the interactions between the variables and imagine that we have an empty graph, such that all variables are independent. 
When assuming an empty graph, we obtain the well-known result that the product of the marginal probabilities is equal to the joint probability. In the graph without interactions, the empty graph, the joint probability for any subset $C\subseteq V$ because of independence is
\begin{align}
p_{C}(x_{C})=\frac{1}{Z_{C}}\exp\left( \sum_{i\in C} \theta_{i}x_{i}\right) = \prod_{i\in C} \frac{1}{Z_{i}}\exp(\theta_{i}x_{i})=\prod_{i\in C}p_{i}(x_{i})
\end{align}
where the normalising constant is $Z_{i} = 1 +\exp(\theta_{i})$. Hence, a quite simple approximation, called inner approximation, is obtained by 
\begin{align}
Z_{C}=\prod_{i\in C}Z_{i}=\prod_{i\in C}(1+\exp(\theta_{i}))
\end{align}
In the inner approximation the interaction parameters are completely ignored in the normalising constant. Hence, the estimate of the normalising constant $Z_{C}$ for subset $C$ will be lower than in reality, and is therefore called a lower bound \citep{Wainwright:2008}.

Another approach is based on the idea that the graph has no cycles and hence is a tree. This idea also underlies the so-called Bethe approximation \citep{Wainwright:2008}. A tree has no cycles and so there are only pairwise connections between the nodes, i.e., the maximal clique size is two. This implies that normalising constant can be written as the product of no more than two nodes each, since those are the cliques. And so, for the Ising model we obtain the normalising constant
$Z=\prod_{(i,j)\in E}Z_{ij}$, where
\begin{align*}
Z_{ij}=1+\exp(\theta_{i})+\exp(\theta_{j})+\exp(\theta_{i}+\theta_{j}+\theta_{ij})
\end{align*}
It is clear that this is computationally much easier with $O(m^{2})$, where $m=|E|$ is the number of edges of $G$, than when  cliques are involved. But in general, for graphs with cycles we obtain an approximation to the true normalising constant that depends on how close the true graph is to a tree \citep{Wainwright:2008}.

%
\subsection{The Curie-Weiss graph}\label{sec:curie-weiss}
By the factorisation property in (\ref{eq:factorisation}) we require normalising with at most the size of the clique $k$ variables. Because the factorisation property is defined over cliques, in which all nodes are connected to each other,  we can invoke the normalising constant for each part of a complete graph, where all nodes are each others neighbour. When within a clique the threshold parameters are all equal and the interaction parameters are all equal, this subgraph can be modeled by a Curie-Weiss graph \citep{Baxter:2007}, and the normalising constant of the Curie-Weiss graph is easier to determine (Marsman, 2018). 

We simplify the model here by letting all interactions between nodes have the same parameter $\theta_{1}$, and the threshold has parameter $\theta_{0}$. Let $\nu$ be the average number of neighbours in $G$. Then the effect of $\nu$ neighbours on any of the nodes is 
\begin{align*}
\theta_{0} + \theta_{1}\sum_{j=1}^{\nu} x_{j}
\end{align*}
In the mean field model we consider the effect on a node as if all nodes were connected to each other and we use the average effect of all $n-1$ other nodes on any one of them. So we obtain \citep{Baxter:2007}
\begin{align*}
\theta_{0} + \theta_{1}\frac{\nu}{n-1}s_{n}
\end{align*}
where $s_{n}=\sum_{j=1}^{n} x_{j}$. Each sum $s_{n}$ can be obtained in $\binom{n}{r}$ ways for $s_{n}=r$. This leads to the probability of a configuration with sum $s_{n}=r$
\begin{align*}
p(x;r)=\frac{1}{Z_{CW}}\binom{n}{r}\exp\left( \theta_{0}r + \frac{\nu}{2(n-1)}\theta_{1}r(r-1) \right) 
\end{align*}
with normalising constant
\begin{align}\label{eq:curie-weiss-constant}
Z_{CW}=\sum_{r=0}^{n}\binom{n}{r}\exp\left( \theta_{0}r+  \frac{\nu}{2(n-1)}\theta_{1} r(r-1) \right)
\end{align}
This version of a complete network as an approximation to one with on average $\nu$ neighbours is sometimes referred to as the Curie-Weiss model. The complexity of the normalising constant $Z_{CW}$ is $O(n)$, and so much smaller than for a graph with any edge distribution, which is $O(2^{n})$ in general. Using the Curie-Weiss version thus makes the problem of determining the normalising constant in the cliques linear and hence scalable. 

We cannot blindly apply the Curie-Weiss computation to the conditional distribution in the Ising model because we have to deal with the values of the variables in the conditioning set. 

We continue with the example of the Ising model corresponding to Figure \ref{fig:cond-vs-do}(a) and determine the Curie-Weiss version of the normalising constant. There is no external field, so $\theta_{0}=0$. For each clique where we need the normalising constant, we take the average of the parameters for interactions. For the clique $C=\{2,3,4\}$ we obtain the average parameter $\theta_{1}=\text{ave}(\theta_{ij}, i,j\in C)$. We obtain two versions of the normalising constant, depending on the value of $x_{2}^{\star}$ being 0 or 1. Considering the normalising constant for the part $p(x_{3},x_{4}\mid x_{2}^{\star})$ we see that the thresholds of $x_{3}$ and $x_{4}$ have changed to $\theta_{3}^{\star}=\theta_{3}+\theta_{23}x_{2}^{\star}$ and $\theta_{4}^{\star}=\theta_{4}+\theta_{24}x_{2}^{\star}$ for $x_{4}$. In this example we had $\theta_{3}=\theta_{4}=0$, and so a threshold appears by conditioning. Therefore, whenever $x_{2}^{\star}$ is 0, then the thresholds remain as if $x_{2}$ was not there, and if $x_{2}^{\star}$ is 1, then the thresholds change to $\theta_{3}'$ and $\theta_{4}'$. In the Curie-Weiss version of the normalising constant we therefore get the average interaction parameter $\theta_{1}=\text{ave}(\theta_{23},\theta_{24},\theta_{34})$ and the average threshold parameter 
\begin{align*}
\theta_{0}^{\star} = 
\begin{cases}
(\theta_{23} +\theta_{24})/2	&\text{ if } x_{2}^{\star}=1\\
0						&\text{ otherwise }
\end{cases}
\end{align*}
With these parameters we fill in equation (\ref{eq:curie-weiss-constant}) to obtain the normalising constant $Z_{CW}$ for the probability $p(x_{3},x_{4}\mid x_{2}^{\star})$ for the clique with $x_{3}$ and $x_{4}$ conditioned on $x_{2}^{\star}$.

From this example we can determine the general rule to obtain the normalising constant for any clique in the factorisation in the conditional distribution. It is clear that for any $j\in A\cap C$ for some clique set $C\in \mathcal{C}$  from the Markov distribution and conditioning set $A\subset V$, the interaction parameters for the $x_{j}^{\star}$ that equals 1 will be added to the threshold parameter, otherwise the threshold parameters  in the clique remain the same. Hence, for clique $C\in \mathcal{C}$ and conditioning set $A$ such that $A\cap C\ne \varnothing$
\begin{align}\label{eq:cw-parameters}
\theta_{0}^{\star}=
\begin{cases}
 \text{ave}(\theta_{i}+\sum_{\substack{j\in A\cap C\\ x_{j}^{\star}=1}}\theta_{ij}, i\in C\cap A^{c}) 	&\text{ if there is $j\in A\cap C$ s.t. } x_{j}^{\star}=1\\
 \text{ave}(\theta_{i},i\in C\cap A^{c})				&\text{ otherwise } 
\end{cases}
\end{align}
and $\theta^{\star}_{1}=\text{ave}(\theta_{ij},i,j\in C)$. 
This simple rule where we change the threshold parameters and leave the interaction parameters allows us tho apply the Curie-Weiss normalisation constant $Z_{CW}$ for each clique in the factorisation of the distribution. In the case that all threshold parameters are equal and all interaction parameters are equal within a clique, then this result is exact. 
\begin{proposition}{\em (Exact normalisation)}\label{prop:curie-weiss-exact}
Let $G$ be a graph induced by the Ising model with cliques $C$ in the set of all cliques $\mathcal{C}$, where for each clique $C$ the threshold parameters are equal and the interaction parameters are equal within the clique of the Ising model. Then for each clique the Curie-Weiss normalisation constant $Z_{C}$ is identical to the exact normalisation constant, and hence, the normalising constant of graph $G$, $Z_{CW}$, is identical to the exact normalisation constant. 
\end{proposition}
We consider the example from Figure \ref{fig:cond-vs-do}(b), where we look at clique $C_{3}=\{2,3,4\}$ (see Figure \ref{fig:cond-vs-do}(a)) and we use $x_{2}^{\star}$ to condition on. If for the clique $C_{3}$ we take all parameters $\theta_{ij}=1$ and, as before, $\theta_{i}=0$, then, we obtain the equivalence according to Proposition \ref{prop:curie-weiss-exact}
\begin{align*}
p_{C_{3}}(x_{C_{3}}\mid\mid x_{2}^{\star})=p_{\{3,4\}}(x_{\{3,4\}}\mid x_{2}^{\star}) = 
\frac{1}{Z_{\{3,4\}\mid 2^{\star}}} \exp( \theta_{23}x_{2}^{\star}x_{3} + \theta_{24}x_{2}^{\star}x_{4}+\theta_{34}x_{3}x_{4} )
\end{align*}
where $Z_{\{3,4\}\mid 2^{\star}}$ is the Curie-Weiss normalisation constant obtained with (\ref{eq:cw-parameters}) and (\ref{eq:curie-weiss-constant}). If $x_{2}^{\star}=1$, then we obtain $Z_{\{3,4\}\mid 2^{\star}}=26.5221$, which equals 
\begin{align*}
1+\exp(\theta_{24})+\exp( \theta_{23})+\exp( \theta_{23} + \theta_{24}+\theta_{34} )
\end{align*}
with each $\theta_{ij}=1$. If we change the edges to $\theta_{23}=1.5$, $\theta_{24}=1$, and $\theta_{34}=1$, then $\theta_{0}^{\star}=(1.5+1)/2=1.25$, and $\theta_{1}^{\star}=1$. And when $x_{2}^{\star}=1$, we obtain $Z_{\{3,4\}\mid 2^{\star}}=41.09614$, while the exact value is $41.31542$. We denote the approximation of the normalising constant for clique $C$ using the averages from (\ref{eq:cw-parameters}) by $\bar{Z}_{C}$.

We see from this small example that in general when the threshold and interaction parameters are different, then using the Curie-Weiss graph is an approximation. We should then ask under what circumstances is the error between the exact and approximate version bounded so that it may still be reasonable to use the approximation. 
\subsection{Bounding the error of $Z_{CW}$}\noindent
By assuming that the deviation of the parameters is small (concetration is high) we can guarantee that the error in the ratio of the exact and approximate normalisation constants (using the Curie-Weiss graph) is bounded. We will assume that the parameters are concentrated around the Curie-Weiss values in terms of sub-Gaussian variables with parameter $\sigma/\sqrt{k}$, where $k$ is the size of the clique. This is a strong assumption. For instance, when the distribution is normal the standard deviation of the distribution of the parameters around $\theta_{0}$ and $\theta_{1}$ is divided by $\sqrt{k}$.

A sub-Gaussian random variable $X$ with mean $0$ is one for which $\sigma>0$ exists such that $\E(\exp(sX))\le \exp(\sigma^{2}s^{2}/2)$ for any $s\in \mathbb{R}$. Taking the threshold and interaction parameters from a sub-Gaussian distribution with parameters $\sigma/k$ and $\sigma/\sqrt{k}$, respectively, together with the Hoeffding bound (see the Appendix and 
\citet[e.g., ][]{Boucheron:2013} or \citet{Venkatesh:2013}) give the approximations 
\begin{align*}
\bar{\theta}_{0} =\theta_{0}+O_{p}(k^{-3/2})\quad\text{and}\quad  
	\bar{\theta}_{1} =\theta_{1}+O_{p}(k^{-3/2})
\end{align*}
where $X_{k}=O_{p}(c_{k})$ means that there exists $K_{\varepsilon}>0$ such that for any $\varepsilon$, $\Prob(|X_{k}|/c_{k}\le K_{\varepsilon})\ge 1-\varepsilon$ for any $k$. Plugging these approximations into the maximal term of the approximate version $\bar{Z}_{C}$ gives a bound on the ratio $\bar{Z}_{C}/Z_{C}$, which converges to 1. 
\begin{proposition}{\em (Error bound on $Z$)} \label{prop:error-bound-zwc}
Let $G$ be a graph associated with a set of random variables $X_{i}$ generated by the Ising probability (\ref{eq:ising}). Furthermore, assume for each clique $C$ in $G$ that the threshold parameters $\theta_{i}$ with mean $\theta_{0}$ and interaction parameters $\theta_{ij}$ with mean $\theta_{1}$, are independent and sub-Gaussian variables with parameter $\sigma/k$ and $\sigma/\sqrt{k}$, respectively.  Then for clique $C$ of size $k$, with probability $1-\delta$, the ratio is
\begin{align}\label{eq:z-ratio}
\frac{\bar{Z}_{C}}{Z_{C}} = \sigma\sqrt{\log(2/\delta)/2}\exp(2k^{-1/2}) \to 1
\end{align}
as $k$ increases.
\end{proposition}
Equivalently, we could say that the difference between the normalising constants is 
\begin{align*}
\bar{Z}_{C}-Z_{C} = Z_{C}\left( 1 +\exp(O_{p}(k^{-1/2})) \right)
\end{align*}
It follows from (\ref{eq:z-ratio}) that when the assumption of the sub-Gaussian parameter $\sigma$ does not hold, then the error of the Curie-Weiss approximation when the interaction parameters are not equal, leads to undesirably large errors. And the constraint for the error of the Curie-Weiss approximation to disappear implies that the deviations of the parameters in the clique cannot be too far from their means. This is mostly problematic for large cliques, but for bounded cliques size, the bound in Proposition \ref{prop:error-bound-zwc} indicates that differences will not be too severe. 

\section{Numerical illustration}\label{sec:numerical-illustration}\noindent
To obtain a clear picture of realistic situations where we require the (conditional) clique normalising constant, we determine the error $\bar{Z}_{C}-Z_{C}$ for different sized cliques and for different values of $\sigma$, the sub-Gaussian parameter.  We vary the clique size from 10 to 100, where the error should be small at clique size 100. We vary the sub-Gaussian parameter $\sigma$ from 1 to 10, where the error is highest at 10. The approximations are computed 100 times.
\begin{figure}[t]\centering
\begin{tabular}{c @{\hspace{3em}} c}
	\includegraphics[width=0.5\textwidth]{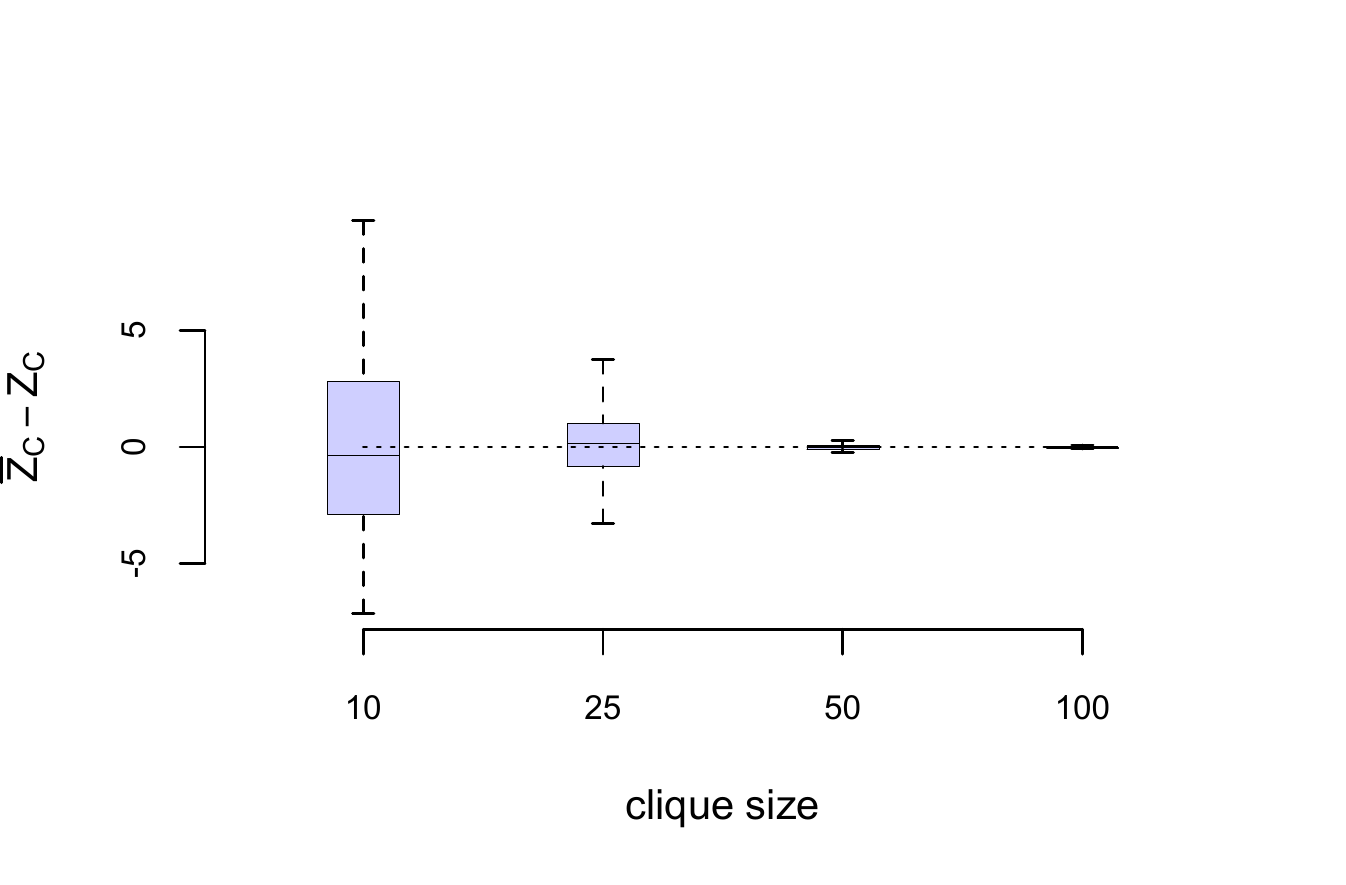} &
	\includegraphics[width=0.5\textwidth]{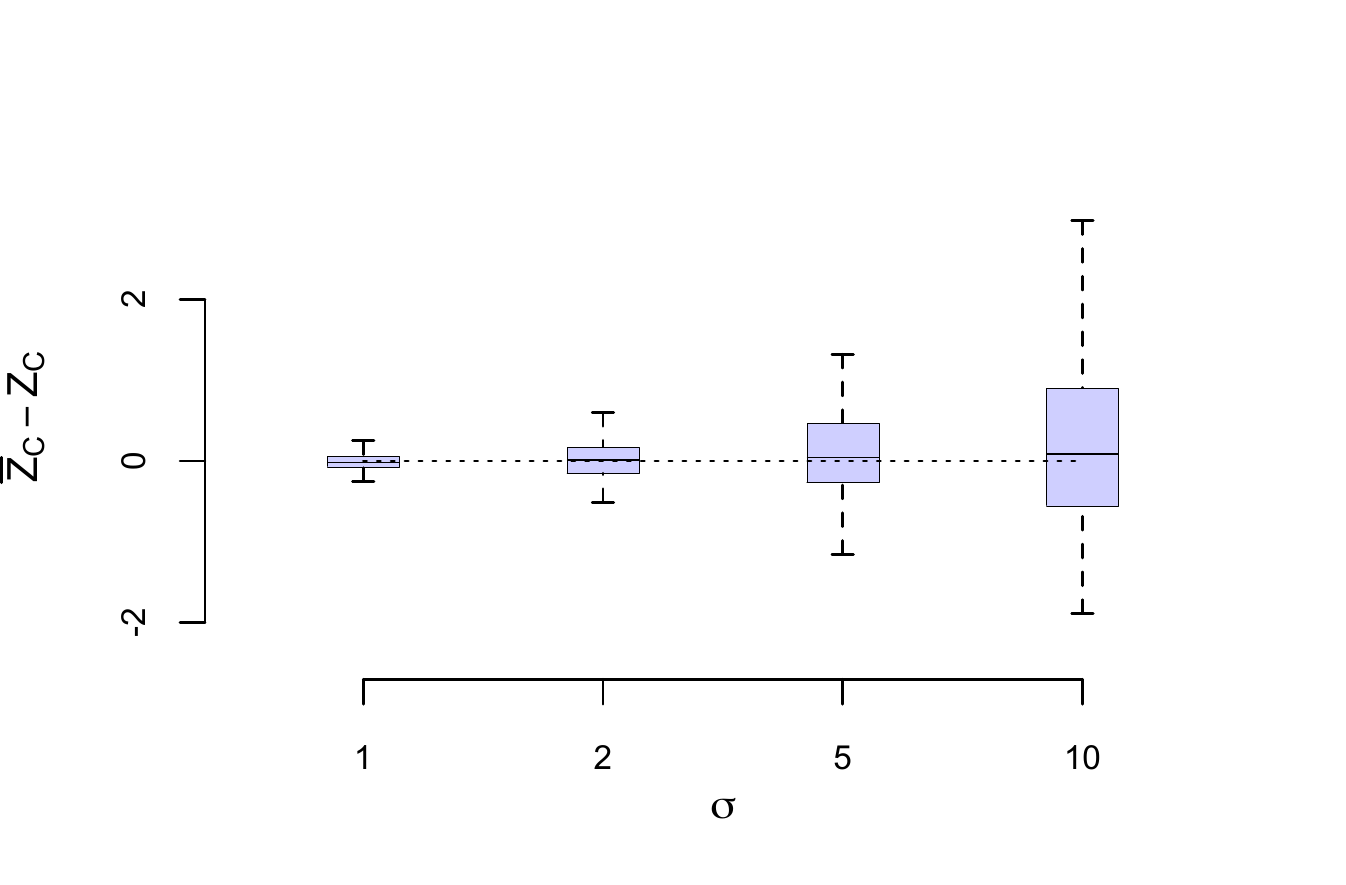}\\
	(a) & (b)
\end{tabular}
\caption{Error of the normalisation constant $\bar{Z}_{C}$ compared to the exact value where all interaction parameters are equal. In (a) the error as a function of clique size $|C|$ and in (b) the error with a network of 50 nodes as a function of the sub-Gaussian parameter $\sigma$.}
\label{fig:approximation-error}
\end{figure}
Figure \ref{fig:approximation-error}(a) shows the error $\bar{Z}_{C}-Z_{C}$ between the Curie-Weiss approximation and the exact value for different clique sizes. As expected, for small clique sizes the error is non-negligible and will have some effect on the probabilities. Note that the probability can either increase or decrease depending on over- or under-estimation of the normalising constant. A similar picture is obtained from Figure \ref{fig:approximation-error}(b), where increasing the parameter $\sigma$ causes larger errors of approximation for the Curie-Weiss normalising constant. Note that single computations of the Curie-Weiss normalising constant can be different for small clique sizes, but the average of several computations (here 100) appears quite accurate. Since computation of the Curie-Weiss normalising constant is fast for each clique, one might consider several estimates to obtain more accurate approximations.

\section{Discussion}\noindent
We considered the issue of intervening in Ising graphs with binary 0-1 nodes, where an intervention was defined by replacement (do-operator). This lead to the fact that interventions can be seen as conditioning on nodes in Ising graphs. To obtain the probabilities in intervention graphs was showed that for graphs with equal connectivities in each clique there is an exact solution to determining the normalisation constant by using the Curie-Weiss model. This simplifies computations considerably, going from $O(2^{n})$ to $O(|\mathcal{C}|k)$, where $n$ is the number of nodes in the graph, $\mathcal{C}$ is the set of all cliques and $k$ is the largest clique size. We also showed that if the connectivities of the edges in the Ising graph are unequal, but the variation is sub-Gaussian, then the error is exponentially small. We confirmed these results with simulations. The simulations indicated that violating the the requirement of a sub-Gaussian variation in connectivities can be diminished if the computations are repeated several times. 
%

\section*{Appendix}\noindent
\begin{proof}{\bf of Proposition \ref{prop:curie-weiss-exact}}
By assumption, for each clique $C$ the threshold parameters $\theta_{j}=\theta_{0}$ are equal, and the interaction parameters $\theta_{jk}=\theta_{1}$ are all equal. Hence, (\ref{eq:cw-parameters}) gives exactly $\theta_{0}$ and $\theta_{1}$. Since the clique is then equal to the Curie-Weiss graph, we obtain the exact normalising constant for the clique. By the factorisation (\ref{eq:factorisation}) each of the terms are conditionally independent, and hence, the product of $\prod_{C\in \mathcal{C}}Z_{C}$ will equal $Z_{CW}$.
\end{proof}
%

\begin{proof}{\bf of Proposition \ref{prop:error-bound-zwc}}
We compare the normalising constant of the exact Curie-Weiss version where $\theta_{i}=\theta_{0}$ for all $i$ and $\theta_{ij}=\theta_{1}$ for all $i\ne j$, and the approximate version, where the $\theta_{i}$ and $\theta_{ij}$ can all be different from the Curie-Weiss parameters. The exact Curie-Weiss version for clique $C$ of size $|C|=k$ we have the normalising constant
\begin{align*}
Z_{C} =\sum_{r=0}^{k}\binom{k}{r}\exp\left( \theta_{0}r+  \frac{\nu}{2(k-1)}\theta_{1} r(r-1) \right)
\end{align*}
The approximate version is the same except that instead of $\theta_{0}$ and $\theta_{1}$ we use the averaged values $\bar{\theta}_{0}$ and $\bar{\theta}_{1}$ of clique $C$ defined in (\ref{eq:cw-parameters}), and we denote this approximate normalising constant for clique $C$ by $\bar{Z}_{C}$.

We consider the maximal value of the sum in the normalising constant in the exact case, that is,
\begin{align*}
\exp\left(\theta_{0}k+\theta_{1}\frac{\nu}{2(k-1)}k(k-1) \right)=\exp(\theta_{0}k+\theta_{1}k\nu/2)
\end{align*}
Then we can use the Hoeffding bound to obtain an approximation to these Curie-Weiss parameters when the values for $\theta_{i}$ and $\theta_{ij}$ are independently obtained from a sub-Gaussian distribution with means $\theta_{0}$ and $\theta_{1}$, respectively. 
For independent sub-Gaussian random variables $X_{i}$ with mean $\mu$ and the same parameter $\sigma$, we obtain the Hoeffding bound for any $t>0$ 
\begin{align}\label{eq:hoeffding-bound}
\Prob \left(  |\bar{X}-\mu | > t \right) \le 2\exp\left( -\frac{2kt^{2}}{\sigma^{2} }\right)
\end{align}
\citep[see, e.g., ][]{Boucheron:2013,Venkatesh:2013}. 
We use the Hoeffding bound and the assumption of sub-Gaussian variables with parameter $\sigma/k$ for the threshold and $\sigma/\sqrt{k}$ for the interaction parameters. Let $\delta=2\exp(-2k^{2}(k-1) t^{2}/\sigma)$, which is the right hand side of the Hoeffding bound with parameter $\sigma/\sqrt{k}$ for the interaction parameters. Then we obtain $t=\sigma\sqrt{\log(2/\delta)/2k^{2}(k-1)}$. For $\theta_{0}$ and $\theta_{1}$ we have from the Hoeffding bound with probability $1-\delta$
\begin{align*}
|\bar{\theta}_{0}-\theta_{0}| \le \sigma\sqrt{\frac{\log(2/\delta)}{2k^{2}(k-1)}} \quad \text{and} \quad 
	|\bar{\theta}_{1}-\theta_{1}| \le \sigma\sqrt{\frac{\log(2/\delta)}{2k^{2}(k-1)}}	
\end{align*}
And so we obtain the approximations
\begin{align*}
\bar{\theta}_{0} =\theta_{0}+O_{p}(k^{-3/2})\quad\text{and}\quad  
	\bar{\theta}_{1} =\theta_{1}+O_{p}(k^{-3/2})
\end{align*}
Plugging these approximations into the maximal term of the approximate version $\bar{Z}_{C}$ gives 
\begin{align*}
\exp\left(\bar{\theta}_{0}k+\bar{\theta}_{1}\frac{\nu}{2(k-1)}k(k-1) \right)=\exp(\theta_{0}k+\theta_{1}k\nu/2 +O_{p}(k^{-1/2}))
\end{align*}
So, the maximal error that we incur for each term is $\exp(O_{p}(k^{-1/2}))$, which converges to 1. Taking this common term out of the sum, shows that we obtain the result. \end{proof}
%


\end{document}